

\input harvmac

\overfullrule=0pt


\def\D{{\scriptscriptstyle D}}

\def\I{{\scriptscriptstyle I}}
\def\J{{\scriptscriptstyle J}}

\def\Q{{\scriptscriptstyle Q}}


\def\CA{{\cal A}}

\def\CH{{\cal H}}


\def\a{\alpha}

\def\d{\delta}

\def\g{\gamma}

\def\th{\theta}


\def\aS{\alpha_s}

\def\bbar{{\overline b}}
\def\Bbar{{\overline B}}
\def\Br{{\rm Br}}
\def\cbar{{\overline c}}
\def\ccdot{\hbox{\kern-.1em$\cdot$\kern-.1em}}
\def\chicJ{\chi_{c\J}}
\def\chiQJ{\chi_{\Q\J}}
\def\chione{\chi_{\Q1}}
\def\chitwo{\chi_{\Q2}}
\def\chizero{\chi_{\Q0}}
\def\dctwo{\d_{c 2}}
\def\dQJ{\d_{\Q\J}}
\def\dQone{\d_{\Q 1}}
\def\dQtwo{\d_{\Q 2}}
\def\dQthree{\d_{\Q 3}}

\def\dash{{\> \over \>}} 		
\def\Dbar{{\overline D}}

\def\gtap{\raise.3ex\hbox{$>$\kern-.75em\lower1ex\hbox{$\sim$}}}
\def\Jpsi{J/\psi}

\def\ltap{\raise.3ex\hbox{$<$\kern-.75em\lower1ex\hbox{$\sim$}}}

\def\Mc{M_c}
\def\MeV{\> {\rm MeV}}

\def\MQ{M_\Q}

\def\onePone{{^1\hskip-0.2 em P_1}}
\def\onethreePJprime{{^{1,3}\hskip-0.15 em P_{J'}}}
\def\onethreeSJprime{{^{1,3}\hskip-0.15 em S_{J'}}}

\def\pbar{{\overline{p}}}

\def\pperp{p_\perp}
\def\psiQ{\psi_\Q}

\def\qbar{{\overline q}}
\def\Qbar{{\overline Q}}

\def\sp{\>\>}

\def\threeDJsinglet{{^3\hskip-0.15 em D_J^{(1)}}}
\def\threePJprime{{^3\hskip-0.15 em P_{\J'}}}
\def\threePJsinglet{{^3\hskip-0.15 em P_{\J}^{(1)}}}
\def\threeSoneoctet{{^3\hskip-0.15 em S_1^{(8)}}}


\def\ninth{{1 \over 9}}

\def\sixth{{ 1\over 6}}

\def\twoninths{{2 \over 9}}


\newdimen\pmboffset
\pmboffset 0.022em
\def\oldpmb#1{\setbox0=\hbox{#1}%
 \copy0\kern-\wd0 \kern\pmboffset\raise
 1.732\pmboffset\copy0\kern-\wd0 \kern\pmboffset\box0}


%
%
\def\appendix#1#2{\global\meqno=1\global\subsecno=0\xdef\secsym{\hbox{#1.}}
\bigbreak\bigskip\noindent{\bf Appendix. #2}\message{(#1. #2)}
\writetoca{Appendix {#1.} {#2}}\par\nobreak\medskip\nobreak}


\nref\HQETreviews{For HQET reviews, see
  M. Wise, ``New Symmetries of the Strong Interactions'', in Proc. of the 6th
  Lake Louise Winter Institute, edited by B.A. Campbell, A.N. Kamall, P.
  Kitching and F.C. Khanna (World Scientific, Singapore, 1991);
  H. Georgi, ``Heavy Quark Effective Theory'', in Proc. of the Theoretical
  Advanced Study Institute (TASI) 1991, edited by R.K. Ellis, C.T. Hill and
  J.D. Lykken (World Scientific, Singapore, 1992);
  B. Grinstein, ``Lectures on Heavy Quark Effective Theory'', in High Energy
  Phenomenology, Proceedings of the Workshop, Mexico City 1-12 July 1991,
  edited by R. Heurta and M.A. Perez, (World Scientific, Singapore);
  M. Neubert, Phys. Rep. {\bf 245} (1994) 259.}
\nref\IsgurWise{N. Isgur and M.B. Wise, Phys. Rev. Lett. {\bf 66} (1991)
 1130.}
\nref\BraatenYuanI{E. Braaten and T.C. Yuan, Phys. Rev. Lett. {\bf 71} (1993)
 1673.}
\nref\BCY{E. Braaten, K. Cheung and T.C. Yuan, Phys. Rev. {\bf D48} (1993)
 4230.}
\nref\Chen{Y.-Q. Chen, Phys. Rev. {\bf D48} (1993) 5181.}
\nref\BraatenYuanII{E. Braaten and T.C. Yuan, Phys. Rev. {\bf D50} (1994)
 3176.}
\nref\Falk{A. Falk and M.E. Peskin, Phys. Rev. {\bf D49} (1994) 3320.}
\nref\Edmonds{A.R. Edmonds, {\it Angular Momentum in Quantum Mechanics}
 (Princeton University Press, Princeton, NJ 1957).}
\nref\PDB{Review of Particle Properties, Phys. Rev. {\bf D50}, Part I
 (1994).}
\nref\Armstrong{T.A. Armstrong {\it et al.} (E760 Collaboration), Phys. Rev.
 Lett. {\bf 69} (1992) 2337.}
\nref\Antoniazzi{L. Antoniazzi {\it et al.} (E705 Collaboration), Phys. Rev.
 {\bf D50} (1994) 4258.}
\nref\CrystalBall{C. Edwards {\it et al.} (Crystal Ball Collaboration),
 Phys. Rev. Lett. {\bf 48} (1982) 70.}
\nref\Bodwin{G.T. Bodwin, E. Braaten and G.P. Lepage, ANL-HEP-PR-94-24
 (1994), unpublished and references therein.}
\nref\Lepage{G.P. Lepage {\it et al.}, Phys. Rev. {\bf D46} (1992) 4052.}
\nref\BDFM{E. Braaten, M. A. Doncheski, S. Fleming and M. L. Mangano,
 Phys. Lett. {\bf B333} (1994) 548.}
\nref\Cacciari{M. Cacciari and M. Greco, Phys. Rev. Lett. {\bf 73} (1994)
 1586.}
\nref\Roy{D.P. Roy and K. Sridhar, CERN-TH.7329/94 (1994), unpublished.}
\nref\Cho{P. Cho, S. Trivedi and M. Wise, CALT-68-1943 (1994), unpublished.}
\nref\Eichten{E. Eichten and F. Feinberg, Phys. Rev. {\bf D23} (1981) 2724.}
\nref\Buchmuller{W. Buchm\"uller, Phys. Lett. {\bf B112} (1982) 479.}
\nref\Isgur{S. Godfrey and N. Isgur, Phys. Rev. {\bf D32} (1985) 189.}
\nref\ChoWise{P. Cho and M. Wise, CALT-68-1954 (1994), unpublished.}
\nref\CDF{CDF Collaboration, Fermilab-conf-94/136-E (1994), unpublished.}
\nref\Close{F.E. Close, RAL-94-093 (1994), unpublished.}
\nref\Braaten{E. Braaten, private communication.}


\nfig\chiQJgraphs{Feynman diagrams which mediate gluon fragmentation to
P-wave $\chiQJ$ orthoquarkonia at order (a) $v^5 (\aS(2 \MQ)/\pi)^2 $ and
(b) $v^5 \aS(2 \MQ)/\pi$.}
\nfig\deltapsigraphs{Feynman diagrams which mediate gluon fragmentation to
D-wave $\dQJ$ orthoquarkonia at order (a) $v^7 (\aS(2\MQ)/\pi)^3$,
(b) $v^7 (\aS(2\MQ)/\pi)^2$ and (c) $v^7 \aS(2\MQ)/\pi$.  These same diagrams
also mediate gluon fragmentation to S-wave $\psi_\Q$ states at order
(a) $v^3 (\aS(2\MQ)/\pi)^3$, (b) $v^7 (\aS(2\MQ)/\pi)^2$ and
(c) $v^7 \aS(2\MQ)/\pi$.  Crossed graphs in (a) and (b) are not displayed.}
%


\def\CITTitle#1#2#3{\nopagenumbers\abstractfont
\hsize=\hstitle\rightline{#1}
\vskip 0.4in\centerline{\titlefont #2} \centerline{\titlefont #3}
\abstractfont\vskip .4in\pageno=0}

\CITTitle{{\baselineskip=12pt plus 1pt minus 1pt
  \vbox{\hbox{CALT-68-1962}\hbox{DOE RESEARCH AND}\hbox{DEVELOPMENT
  REPORT}}}}
{Spin Symmetry Predictions}{for Heavy Quarkonia Alignment}
\centerline{
  Peter Cho\footnote{$^1$}{Work supported in part by a DuBridge Fellowship and
  by the U.S. Dept. of Energy under DOE Grant no. DE-FG03-92-ER40701.}
  and Mark B. Wise\footnote{$^2$}{Work supported in part by
  the U.S. Dept. of Energy under DOE Grant no. DE-FG03-92-ER40701.}}
\centerline{Lauritsen Laboratory}
\centerline{California Institute of Technology}
\centerline{Pasadena, CA  91125}

\vskip .2in
\centerline{\bf Abstract}
\bigskip

	We investigate the implications of spin symmetry for heavy quarkonia
production and decay.  We first compare spin symmetry predictions for
charmonia and bottomonia radiative transitions with data and find they agree
quite well.  We next use spin symmetry along with nonrelativistic QCD power
counting to determine the leading order alignment of P and D-wave
orthoquarkonia that are produced through gluon fragmentation.  Finally, we
discuss a mechanism which may resolve the current factor of 30 discrepancy
between theoretical predictions and experimental measurements of prompt $\psi'$
production at the Tevatron.  This mechanism involves gluon fragmentation to a
subleading Fock component in the $\psi'$ wavefunction and yields 100\%
transversely aligned $\psi'$'s to lowest order.  Observation of such a large
alignment would provide strong support for this resolution to the $\psi'$
problem.

\Date{11/94}

\newsec{Introduction}

	Spin symmetry has played an important role in the study of
heavy hadrons during the past several years.  Many of its consequences
for mesons and baryons with quark content $Q\qbar$ and $Qqq$ have been
investigated within the context of the Heavy Quark Effective Theory
(HQET) \HQETreviews.  The spin and flavor symmetries of QCD that
become exact in the $\MQ \to \infty$ limit are transparent in this
effective theory framework.  But even without the HQET apparatus,
several aspects of charm and bottom hadron phenomenology may be
understood from symmetry considerations alone.  For example, the rates
for strong and electromagnetic transitions between states belonging to
separate spin doublets are simply related by group theory factors.
Moreover, the sums of partial widths for each member of a doublet to
decay to states in another doublet are equal.  These predictions that
follow from spin symmetry in the infinite mass limit appear to be
reasonably well satisfied for finite mass hadrons in the charm and
bottom sectors \IsgurWise.

	In this letter, we consider the implications of spin symmetry
for heavy quarkonia.  Since the spins of both the quark and antiquark
inside a $Q\Qbar$ bound state are separately conserved as $\MQ \to
\infty$, transitions between members of different quarkonia multiplets
within the same flavor sector are more tightly constrained than their
heavy-light meson analogues.  Ratios of decay rates can be predicted
and checked against the large body of charmonia and bottomonia data
collected over the past two decades.  Spin symmetry also constrains
$Q\Qbar$ bound state formation and ties in closely with recent
perturbative QCD calculations of parton fragmentation to heavy
quarkonia \refs{\BraatenYuanI{--}\BraatenYuanII}.  As we shall see,
the union of these ideas yields several interesting predictions which
can be experimentally tested at hadron colliders.

	Our paper is organized as follows.  We first review the
general relation that spin symmetry imposes upon heavy hadron
transitions in section 2.  We then test this relation for quarkonia by
comparing its predictions for $Q\Qbar$ electromagnetic transitions
with experimental data.  In section 3, we use the spin symmetry
relation along with nonrelativistic QCD power counting arguments to
determine the leading order alignment of P and D-wave orthoquarkonia
produced via gluon fragmentation.  Finally, we discuss the
implications of spin symmetry for a fragmentation mechanism which may
resolve the current large discrepancy between theory and experiment in
prompt $\psi'$ production at the Tevatron.

\newsec{Radiative Quarkonia Transitions}

	It is useful to recall the basic symmetry relation which
constrains transitions between any initial and final hadronic states
$H$ and $H'$ that contain heavy constituents \IsgurWise.  For $Q\qbar$
mesons and $Qqq$ baryons, we let $j^{(\prime)}_h$, $j^{(\prime)}_\ell$
and $J^{(\prime)}$ denote the angular momentum of $H^{(\prime)}$'s
heavy, light and total degrees of freedom.  As the mass of the heavy
constituents inside $H^{(\prime)}$ tends to infinity, the separate
angular momenta of the light and heavy degrees of freedom become good
quantum numbers.  A similar phenomenon occurs for $Q\Qbar$ quarkonia.
But in this case, the quantum number $j^{(\prime)}_h$ represents just
the total spin of the $Q\Qbar$ pair, while $j^{(\prime)}_\ell$ stands
for the bound state's remaining angular momentum which includes the
orbital contribution from the heavy quark and antiquark.

	In the infinite mass limit, the amplitude for a general $H \to H'$
process that is mediated by an interaction Hamiltonian $\CH_\I$
and results in the emission or absorption of some light quanta can be
decomposed as
\eqn\preamp{\eqalign{
i\CA \sp = \sp & \langle J' J'_z | \CH_\I | J J_z \rangle \cr
& = \sum \langle J' J'_z | j'_\ell j'_{\ell z} ; j'_h j'_{h z} \rangle
\langle j'_\ell j'_{\ell z}; j'_h j'_{h z}| \CH_\I | j_\ell j_{\ell z};
j_h j_{h z} \rangle \langle j_\ell j_{\ell z} ; j_h j_{h z} | J J_z \rangle \cr
& =  \sum \langle J' J'_z | j'_\ell j'_{\ell z} ; j'_h j'_{h z} \rangle
\langle j'_\ell j'_{\ell z} | \CH_\I | j_\ell j_{\ell z} \rangle
\langle j_\ell j_{\ell z} ; j_h j_{h z} | J J_z \rangle
\d_{j^{}_h j'_h} \d_{j^{}_{h z} j'_{h_z}} \cr}}
where repeated angular momentum labels are summed.  If the reaction
proceeds through the $L$-th multipole of $\CH_\I$, the amplitude
becomes proportional to the reduced matrix element $\langle j'_\ell ||
L || j_\ell \rangle$ and may be conveniently rewritten in terms of a
6$j$-symbol \refs{\Falk,\Edmonds}:
\eqn\amp{\eqalign{i \CA &=
\Bigl[ \sum \langle J' J'_z | j'_\ell j'_{\ell z} ; j_h j_{h z} \rangle
\langle L L_z ; j'_\ell j'_{\ell z} | j_\ell j_{\ell z} \rangle
\langle j_\ell j_{\ell z} ; j_h j_{h z} | J J_z \rangle \Bigr]
\langle j'_\ell || L || j_\ell \rangle \cr
&= (-1)^{L+j'_\ell+j_h+J} \sqrt{(2j_\ell+1)(2J'+1)}
\left\{\matrix{L & j'_\ell & j_\ell \cr
	       j_h & J & J'\cr}\right\}
\langle L \, (J_z - J'_z) ; J' J'_z  | J J_z \rangle
\langle j'_\ell || L || j_\ell \rangle. \cr}}
The consequences of spin symmetry follow from this last formula.
\foot{The only $6j$-symbols which we will need for applications of \amp\ in
this letter are
$$ \eqalign{&
\left\{\matrix{L & 0 & L \cr j_h & J & j_h\cr}\right\} =
{(-1)^{L+j_h+J} \over \sqrt{(2L+1)(2j_h+1)}}\cr}
\quad {\rm and} \quad
\eqalign{&
\left\{\matrix{L & L & 0 \cr j_h & j_h & J'\cr}\right\} =
{(-1)^{L+j_h+J'} \over \sqrt{(2L+1)(2j_h+1)}}.\cr}$$}

	We will apply eqn.~\amp\ to heavy quarkonia transitions.  In
order to test its validity, we first investigate charmonia and
bottomonia radiative decays.  We start with the electric dipole
process $^{1,3}P_J \to \onethreeSJprime + \gamma$.  In the initial
state, the heavy spin angular momentum $j_h=0$ or $j_h=1$ can be
combined with the P-wave orbital angular momentum $j_\ell=1$ to form a
$h_\Q$ or $\chi_{\Q \J}$ quarkonium. After the $L=1$ radiative
transition, the final state contains an S-wave $\eta_\Q$ or $\psiQ$
with $j'_\ell=0$.  Eqn.~\amp\ implies that the decay rate
$\Gamma(^{1,3}P_J \to \onethreeSJprime + \gamma) \propto E_\g^3$,
averaged and summed over initial and final polarizations, is
independent of $J$ and $J'$ in the infinite mass limit.  For finite
mass quarkonia, the rate indirectly depends upon these angular
momentum quantum numbers through formally subleading but
phenomenologically important splittings between members of the S and
P-wave multiplets.  Taking these splittings into account, we find the
following radiative partial width ratios in the $c\cbar$ sector:
\eqna\ratiosI
\foot{We use the experimentally measured paracharmonia masses
$M_{\eta_c}=2979 \MeV$ \PDB, $M_{h_c} = 3526 \MeV$
\refs{\Armstrong,\Antoniazzi} and $M_{\eta'_c} = 3592 \MeV$ \CrystalBall\ to
predict the $\Gamma(h_c \to \eta_c + \gamma)$ and $\Gamma(\eta'_c \to h_c
+ \gamma)$ entries in eqns.~\ratiosI{a}\ and (2.6a).}
$$ \eqalignno{
& \Gamma(\chi_{c0} \to \Jpsi + \gamma) : \Gamma(\chi_{c1} \to \Jpsi + \gamma) :
  \Gamma(\chi_{c2} \to \Jpsi + \gamma) : \Gamma(h_c \to \eta_c + \gamma) & \cr
& \quad = 0.095 : 0.20 : 0.27 : 0.44
\>\quad\qquad\qquad\qquad\qquad\qquad\qquad\qquad ({\rm Theory}) &\ratiosI a
\cr
& \quad = 0.092 \pm 0.041 : 0.24 \pm 0.04 : 0.27 \pm 0.03 : {\rm unmeasured}.
\qquad ({\rm Experiment}) & \ratiosI b \cr} $$
Analogous spin symmetry predictions in the $b\bbar$ sector cannot be
checked against data, for absolute radiative partial widths have not
yet been measured.  However, we can compare theoretical and
experimental values for the branching fraction ratio
\eqn\RJ{R_\J = {\Gamma(\chi_{b\J}(2P) \to \Upsilon(1S)+\gamma) \over
\Gamma(\chi_{b\J}(2P) \to \Upsilon(2S)+\gamma)}
\propto \Bigl[ {E_\gamma(2P \to 1S) \over E_\gamma(2P \to 2S)} \Bigr]^3}
as a function of $J$:
\eqna\ratiosII
$$ \eqalignno{
R_0:R_1:R_2 &= 3.22 : 2.55 : 2.28
\>\>\quad\qquad\qquad\qquad\qquad\qquad\qquad ({\rm Theory}) & \ratiosII a \cr
&= 5.11 \pm 4.13 : 2.47 \pm 0.60 : 2.28 \pm 0.47.
\>\qquad\qquad ({\rm Experiment}) & \ratiosII b \cr} $$
We do not include the corresponding prediction for $\Gamma(h_b(2P) \to
\eta_b(2S) + \gamma)/\Gamma(h_b(2P) \to \eta_b(1S) + \gamma)$ in
eqn.~\ratiosII{a} since the parabottomonium phase space factors are
uncertain as the $\eta_b(1S)$, $\eta_b(2S)$ and $h_b(2P)$ masses have
not been experimentally measured.

	We can further test spin symmetry for heavy quarkonia by
considering $^{1,3}S_\J \to$ $\onethreePJprime + \gamma$ electric
dipole decays.  Inserting $j_\ell = 0$, $j'_\ell=1$ and $L=1$ into
amplitude \amp, we find $\Gamma({}^3S_1 \to \threePJprime + \gamma)
\propto \bigl((2J'+1)/9 \bigr) \> E_\gamma^3$ and
$\Gamma({}^1S_0 \to \onePone + \gamma) \propto E_\gamma^3$.  These partial
width results yield the following charmonium and bottomonium sector ratios:
\eqna\ratiosIII
$$ \eqalignno{
& \Gamma(\psi' \to \chi_{c0} + \gamma) : \Gamma(\psi' \to \chi_{c1} + \gamma) :
\Gamma(\psi' \to \chi_{c2} + \gamma) : \Gamma(\eta'_c \to h_c + \gamma) & \cr
& \quad = 9.3 : 7.9 : 5.4 : 1.3
\quad\qquad\qquad\qquad \qquad\qquad\qquad\qquad ({\rm Theory}) & \ratiosIII
 a \cr
& \quad = 9.3 \pm 0.8 : 8.7 \pm 0.8 : 7.8 \pm 0.8 : {\rm unmeasured}
\sp\quad\qquad ({\rm Experiment}) & \ratiosIII b \cr
&&\cr
\eqna\ratiosIV
& \Gamma(\Upsilon(2S) \to \chi_{b0}(1P) + \gamma) :
\Gamma(\Upsilon(2S) \to \chi_{b1}(1P) + \gamma) :
\Gamma(\Upsilon(2S) \to \chi_{b2}(1P) + \gamma) & \cr
& \quad  = 4.29 : 6.70 : 6.59
\>\quad\qquad\qquad\qquad \qquad\qquad\qquad\qquad ({\rm Theory}) & \ratiosIV
 a \cr
& \quad  = 4.3 \pm 1.0 : 6.7 \pm 0.9 : 6.6 \pm 0.9
\>\quad\qquad\qquad\qquad\qquad ({\rm Experiment}) & \ratiosIV b \cr
&&\cr
\eqna\ratiosV
& \Gamma(\Upsilon(3S) \to \chi_{b0}(2P) + \gamma) :
\Gamma(\Upsilon(2S) \to \chi_{b1}(2P) + \gamma) :
\Gamma(\Upsilon(2S) \to \chi_{b2}(2P) + \gamma) & \cr
& \quad  = 7.0 : 11.3 : 12.3
\>\>\quad\qquad\qquad\qquad \qquad\qquad\qquad\qquad ({\rm Theory}) & \ratiosV
 a \cr
& \quad  = 5.4 \pm 0.6 : 11.3 \pm 0.6 : 11.4 \pm 0.8 .
\>\sp\quad\qquad\qquad\qquad ({\rm Experiment}) & \ratiosV b \cr} $$
Again we do not include parabottomonium partial widths in eqns.~\ratiosIV{a}\
and \ratiosV{a}\ since their symmetry breaking phase space factors are
uncertain.

	Comparing the theoretical and experimental numbers in
eqns.~\ratiosI{}\ - \ratiosV{}, we see that the predictions of spin
symmetry agree quite well with the quarkonia data. The discrepancies
in a given $b\bbar$ mode are smaller than those in its $c\cbar$
analogue, as one would expect.  Spin symmetry also works best for the
lowest lying $Q\Qbar$ states where the ratio of the heavy quarks'
kinetic energy to their rest mass is minimal. But the general success
of these radiative transition findings bolsters one's confidence that
results for other processes which follow from the amplitude expression
in eqn.~\amp\ will also work well.  We therefore use it to investigate
quarkonia production in the following section.

\newsec{Orthoquarkonia Alignment}

	Quarkonia physics involves a number of different scales
separated by the velocity $v$ of the heavy quarks inside $Q\Qbar$
bound states.  These scales are set by the quarks' mass $\MQ$, typical
momentum $\MQ v$ and kinetic energy $\MQ v^2$.  Within the past few
years, an effective field theory formalism called Nonrelativistic
Quantum Chromodynamics (NRQCD) has been established to keep track of
this quarkonia scale hierarchy \Bodwin.  The effective theory is based
upon a double power series expansion in the heavy quark velocity and
strong interaction fine structure constant.  Its power counting rules
allow one to methodically determine the most important contributions
to a particular quarkonium process \Lepage.  We will utilize these
NRQCD rules in conjunction with spin symmetry arguments to investigate
quarkonia production via gluon fragmentation.  As we shall see, these
tools yield phenomenologically interesting predictions for
orthoquarkonia alignment.

	We first consider $g \to \chiQJ$ fragmentation.  The Feynman
diagrams that mediate this process at lowest order in the the velocity
expansion are illustrated in figs.~1a and 1b.  The graphs contribute
to $\chiQJ$ production through scattering collisions like $gg \to gg^*
\to gg \chiQJ$.  Such fragmentation reactions dominate over lower
order parton fusion processes when the lab frame energy $q_0$ of the
fragmenting gluon $g^*$ is large, but its squared four-momentum $q^2$
is close to the bound state's squared mass $M^2_{\chiQJ} \simeq
(2\MQ)^2$.  In this case, $g^*$ is transverse up to corrections of
order $q^2/q_0^2$.  Consequently, the fragmenting gluon is essentially
real so long as its energy $q_0$ is significantly greater than
$M_{\chiQJ}$.

	The $Q\Qbar$ states that emerge from the shaded ovals in
\chiQJgraphs\ reside within the first two Fock components of the
$\chiQJ$ wavefunction
\eqn\chiwavefunc{|\, \chiQJ\rangle = O(1) |\, Q\Qbar\, [ {}^3P_J^{(1)}] \,
 \rangle + O(v) |\, Q\Qbar\, [ {}^3S_1^{(8)}] \,  g \rangle + \cdots.}
The angular momentum quantum numbers of the $Q\Qbar$ pairs are
indicated in spectroscopic notation inside the square brackets, and
their color configurations are labeled by singlet or octet superscripts.  In
\chiQJgraphs a, the incoming hard gluon with four-momentum $q$ fragments into
$Q\Qbar[{}^3P_J^{(1)}]$ plus an outgoing hard gluon.  The rate for this
process is proportional to the square of the first derivative of the P-wave
bound state's wavefunction evaluated at the origin $|\,R_1'(0)|\,^2 \sim v^5$.
In \chiQJgraphs b, the incoming gluon turns into $Q\Qbar[{}^3S_1^{(8)}]$ which
appears in the next-to-leading Fock component in \chiwavefunc.  The formation
rate for the colored S-wave pair is proportional to the square of its
wavefunction $|\,R_8(0)|\,^2 \sim v^3$.  The ${}^3S_1^{(8)}$ state
subsequently converts to a $\chiQJ$ through a soft gluon coupling which is
depicted as a ``black box'' in \chiQJgraphs b.  This ``black box'' may be
regarded as mediating the dominant chromoelectric dipole transition
${}^3S_1^{(8)} \to \threePJsinglet + g$.  Although the resulting P-wave state
has order unity overlap with the full $\chiQJ$ wavefunction, the soft
gluon emission costs one power of $v$ in the amplitude and two in the
rate.  The total transition in \chiQJgraphs b therefore proceeds at
$O(v^5)$.  Alternatively, the ``black box'' may be viewed as combining
the ${}^3S_1^{(8)}$ $Q\Qbar$ pair with a soft gluon to form the
complete $|\,Q\Qbar\, [ {}^3S_1^{(8)}] \, g\rangle$ Fock component in
eqn.~\chiwavefunc\ which has $O(v)$ overlap with $|\, \chiQJ\rangle$.
This interpretation again leads one to conclude that the transition in
\chiQJgraphs b takes place at $O(v^5)$ \BraatenYuanII.

	Although the Feynman diagrams in \chiQJgraphs\ are the same
order in the NRQCD velocity expansion, their dependence upon the short
distance fine structure constant is different.  The graphs in
\chiQJgraphs a involve two hard gluons and are suppressed by
$\aS(2\MQ)/\pi$ relative to the graph in \chiQJgraphs b which has only
one.  This perturbative QCD suppression is partly offset by the
numerical values $H_1 \simeq 15 \MeV$ and $H'_8 \simeq 3 \MeV$ for the
nonperturbative matrix elements associated with the color-singlet and
octet graphs \BraatenYuanII.  However, prompt $\chiQJ$ production at
high transverse momenta is known to be dominated by gluon
fragmentation to $Q\Qbar[{}^3S_1^{(8)}]$ \refs{\BDFM{--}\Roy}.
Therefore, we will simply neglect the subleading contributions to $g \to
\chiQJ$ which come from the diagrams in \chiQJgraphs a.
\foot{In order to completely determine the $O((\aS(2\MQ)/\pi)^2)$
contribution to $g \to \chiQJ$ fragmentation, the interference between
the graph in \chiQJgraphs b and its one-loop corrections must be
calculated along with the diagrams in \chiQJgraphs a.  Such
interference terms have not been included so far in $g \to \chiQJ$
fragmentation function computations.}

	The intermediate ${}^3S_1^{(8)}$ state in \chiQJgraphs b has
the same quantum numbers as the incoming hard gluon.  In particular,
it is transversely aligned at high energies.  Spin symmetry fixes the
degree to which the final $\chiQJ$ inherits the color-octet state's
alignment through the soft gluon processes ${}^3S_1^{(8)} \to
\threePJsinglet + g$ or ${}^3S_1^{(8)} + g \to \threePJsinglet$.  The
populations of all the individual $\chiQJ$ helicity components may be
determined by setting $j_\ell=0$, $j'_\ell=1$ and $L=1$ in the general
amplitude expression \amp.  After a straightforward computation, we
find that the helicity levels are filled according to
\eqn\chiQJpopulations{
\chizero : \chione^{(h=0)} : \chione^{(|h|=1)} :
\chitwo^{(h=0)} : \chitwo^{(|h|=1)} : \chitwo^{(|h|=2)}
= \ninth:\sixth:\sixth:{1 \over 18}:{3 \over 18}:{6 \over 18}.}

	A few points about this result should be noted.  Firstly, if
we compare the helicity populations in \chiQJpopulations\ with their
unaligned counterparts
\eqn\unalignedchiQJpopulations{
\chizero : \chione^{(h=0)} : \chione^{(|h|=1)} :
\chitwo^{(h=0)} : \chitwo^{(|h|=1)} : \chitwo^{(|h|=2)}
= \ninth:\ninth:\twoninths:\ninth:\twoninths:\twoninths,}
we see that gluon fragmentation at high transverse momenta yields a
sizable $\chiQJ$ alignment.  Secondly, the ratios in
eqn.~\chiQJpopulations\ are consistent with the leading color-octet
terms in the polarized $g \to \chicJ$ fragmentation functions which
were calculated in ref.~\Cho.  The effect of the $\chicJ$ alignment
upon the angular distribution of photons that result from the electric
dipole transition $\chicJ \to \Jpsi + \gamma$ as well as the induced
$\Jpsi$ alignment were discussed in this previous article.  We will
not reiterate those findings of ref.~\Cho\ here, but we note that they
may be simply recovered from spin symmetry arguments.  Finally,
uncertainty in the numerical value for the nonperturbative matrix
element $H'_8$ drops out from the $\chiQJ^{(h)}$ ratios in
\chiQJpopulations.  Therefore, they represent more reliable
predictions than corresponding estimates for absolute $g \to
\chiQJ^{(h)}$ fragmentation probabilities which depend upon the
presently poorly constrained $H'_8$ parameter.

	We consider next gluon fragmentation to D-wave quarkonia.  In
the absence of a well-established nomenclature convention, we adopt
the symbol $\dQJ$ to refer to $L=2$ quarkonia with total angular
momentum $J$.  Quark model calculations indicate that $n=1$ and $n=2$
$\d_{b\J}$ bottomonia lie below the $B\Bbar$ threshold.  On the other
hand, all $n=1$ $\d_{c\J}$ charmonia are predicted to have masses
greater than $2 M_\D$ but less than $M_D+M_{D^*}$
\refs{\Eichten{--}\Isgur}.
\foot{Evidence for a 3.836 GeV $1 {}^3D_2$ charmonium state has recently been
reported in ref.~\Antoniazzi.}
The $1 {}^3D_1$ and $1 {}^3D_3$ states may decay to $D\Dbar$ and are
broad.  But parity forbids their $1 {}^1D_2$ and $1{}^3D_2$
counterparts from decaying to two spinless mesons.  The $J=2$
charmonia are consequently quite narrow.  Gluon fragmentation to
$1{}^1D_2$ paracharmonia has been found to yield a nonnegligible
$p\pbar \to 1{}^1D_2+X$ cross section at the Tevatron \ChoWise.
Fragmentation to experimentally more accessible $1{}^3D_2$
orthocharmonia starts at the same order in the velocity expansion but
one lower order in $\aS(2 \Mc)/\pi$.  The $1 {}^3D_2$ states should
therefore be produced in greater abundance.

	Gluon fragmentation to $\dQJ$ orthoquarkonia proceeds at
$O(v^7)$ via Feynman diagrams like those illustrated in figs.~2a, 2b
and 2c.  The states emerging from the shaded ovals in these graphs
correspond to the $Q\Qbar$ pairs in the following Fock components of
the $\dQJ$ wavefunction:
\eqn\Dwavefunc{|\, \dQJ\,\rangle = O(1) |\, Q\Qbar\, [ {}^3D_J^{(1)}] \,\rangle
+ O(v) |\, Q\Qbar\, [ {}^3P_{J'}^{(8)}] \,  g \rangle
+ O(v^2) |\, Q\Qbar\, [ {}^3S_{1}^{(8)}] \,  g g \rangle + \cdots.}
The diagrams in \deltapsigraphs a and 2b are suppressed relative to
that in \deltapsigraphs c by $(\aS(2\MQ)/\pi)^2$ and $\aS(2\MQ)/\pi$
respectively.  So the dominant contribution to $g \to \dQJ$
fragmentation comes from the Fock component that contains the S-wave
color-octet state.  Its subsequent conversion to $\dQJ$ through long
distance processes such as $\threeSoneoctet \to \threeDJsinglet + g +
g$ and $\threeSoneoctet + g + g \to \threeDJsinglet$ conserves
$j_h=1$.  We can thus determine the alignment which the $\dQJ$
orthoquarkonium inherits from its transverse ${}^3S_1^{(8)}$
progenitor by setting $j_\ell=0$, $j'_\ell=2$ and $L=2$ in eqn.~\amp.
We find that the individual $\dQJ^{(h)}$ helicity states are populated
according to
\eqn\Dwavepopulations{\eqalign{
& \dQone^{(h=0)} : \dQone^{(|h|=1)} : \dQtwo^{(h=0)} : \dQtwo^{(|h|=1)} :
\dQtwo^{(|h|=2)} : \dQthree^{(h=0)} : \dQthree^{(|h|=1)} : \dQthree^{(|h|=2)}
: \dQthree^{(|h|=3)} \cr
& \quad = {3 \over 50} : {7 \over 50}
: {3 \over 30} : {5 \over 30} : {2 \over 30} :
{3 \over 75} : {7 \over 75} : {10 \over 75} : {15 \over 75}.}}

	Once $\dQJ$ orthoquarkonia are formed, spin symmetry fixes the
leading order transfer of alignment from these D-wave states to S and
P-wave orthoquarkonia descendants in subsequent strong and electromagnetic
decays.  Consider for example the decay mode $\d_{c2} \to \Jpsi \> \pi \pi$
in the $c\cbar$ sector.  This reaction conserves heavy spin angular momentum
since the outgoing pions are soft.  So we can again use eqn.~\amp\ to compute
the alignment which the final $\Jpsi$ inherits from the initial $\dctwo$:
\eqn\feeddown{\eqalign{
\Jpsi^{(h=0)} &= \Br(\dctwo \to \Jpsi \> \pi\pi)
  \Bigl[0 \> \dctwo^{(h=0)} + {1 \over 6} \> \dctwo^{(|h|=1)}
  + {2 \over 3} \> \dctwo^{(|h|=2)} \Bigr] \cr
\Jpsi^{(|h|=1)} &= \Br(\dctwo \to \Jpsi \> \pi\pi)
  \Bigl[1 \> \dctwo^{(h=0)} + {5 \over 6} \> \dctwo^{(|h|=1)}
  + {1 \over 3} \> \dctwo^{(|h|=2)}\Bigr]. \cr}}
The names of the helicity states in this expression denote
fragmentation probabilities.  Inserting the $J=2$ D-wave results from
eqn.~\Dwavepopulations, we deduce that $\dctwo \to \Jpsi \> \pi\pi$
feeddown produces longitudinal and transverse $\Jpsi$'s in the ratio
$\Jpsi^{(h=0)} : \Jpsi^{(|h|=1)} = 1:3.6$.

	We now come to the most interesting application of spin
symmetry ideas to heavy quarkonia systems.  Specifically, we consider
its implications for S-wave quarkonia formation.  Prompt $\Jpsi$ and
$\psi'$ production at the Tevatron are currently the subject of
significant theoretical and experimental interest.  Previous studies
have found that parton fusion alone underestimates the rate for
$\Jpsi$ production at large transverse momenta by more than a factor
of 10.  But if fragmentation is included, the disagreement between
theory and experiment reduces to roughly a factor of 2
\refs{\BDFM{--}\Roy}.  The remaining discrepancy may be attributed to
a number of theoretical uncertainties in perturbative QCD
fragmentation function computations.  The application of fragmentation
ideas to $\Jpsi$ production thus represents a qualitative success.

	The situation for prompt $\psi'$ production is totally
different.  The prediction for $d\sigma(p\pbar \to \psi'+X)/d\pperp$
at high $\pperp$ underestimates the measured cross section by a factor
of 30 even after fragmentation is taken into account \CDF.  Recently,
gluon fragmentation to some of the $n=2$ $\chi'_{c\J}$ states that lie
above the $D\Dbar$ threshold has been proposed as a possible
significant source of $\psi'$'s \refs{\Cho,\Close}.  This mechanism
requires an uncomfortably large radiative branching fraction
$\Br(\chi'_{c\J} \to \psi'+\gamma) \simeq 5 \dash 10 \%$ in order to explain
the factor of 30 discrepancy between theory and data.  However, a search for
the photons that result from the electromagnetic transition along with an
analysis of their angular distribution would at least provide a means for
testing this proposal.

	There is another possible resolution to the $\psi'$ surplus
problem which does not involve any charmonia states above the $D\Dbar$
threshold \Braaten.  Instead, the answer may lie in fragmentation to higher
Fock state components within the $\psi'$ wavefunction
\eqn\psiwavefunc{\eqalign{
|\, \psi' \rangle &= O(1) |\, c\cbar\,[{}^3S_1^{(1)}] \,\rangle
+ O(v) |\, c\cbar\, [ {}^3P_J^{(8)}] \,  g \rangle \cr
& \quad + O(v^2) |\, c\cbar\, [ {}^3D_J^{(1,8)}] \,  g g \rangle
+ O(v^2) |\, c\cbar\, [ {}^1S_0^{(8)}] \,  g \rangle
+ O(v^2) |\, c\cbar\, [ {}^3S_1^{(1,8)}] \,  g g \rangle +
\cdots.}}
Gluon fragmentation to the $c\cbar[{}^3S_1^{(1)}]$ pair in the leading
Fock component starts at $O(v^3 (\aS(2\Mc)/\pi)^3)$ via Feynman
diagrams like the one illustrated in \deltapsigraphs a.  Fragmentation
to the $c\cbar$ states in all other Fock components of $|\, \psi'\,
\rangle$ takes place at higher order in the velocity expansion.  But
the heavy charm-anticharm pairs in the subleading Fock components can
be formed at lower orders in $\aS(2\Mc)/\pi$.  In particular, the
transformation shown in \deltapsigraphs c of a hard gluon into a
${}^3S_1^{(8)}$ $c\cbar$ pair followed by its conversion into $\psi'$
via the emission or absorption of two soft gluons occurs at $O(v^7
\aS(2\Mc)/\pi)$.  If we evaluate the relative strengths of the $\psi'$
production processes in figs.~2a and 2c by setting $v^2 \simeq 0.3$
and $\aS(2\Mc)/\pi \simeq 9.0 \times 10^{-2}$, we find $v^3
(\aS(2\Mc)/\pi)^3 \simeq 1.2 \times 10^{-4}$ and $v^7 (\aS(2\Mc)/\pi)
\simeq 1.4 \times 10^{-3}$.  Hence, gluon fragmentation to the
color-octet $c\cbar$ pair within the last Fock state of
eqn.~\psiwavefunc\ may naively be an order of magnitude more important
than that to the color-singlet pair within the first Fock state.  If
so, its inclusion could resolve the $\psi'$ problem.
\foot{Gluon fragmentation to the $c\cbar$ pair within the $| c\cbar
[{}^3S_1^{(8)}] g g \rangle$ Fock component of $|\psi\rangle$ also
enhances prompt $\Jpsi$ production.  However, its impact is smaller
than that for $\psi'$ production since high transverse momenta
$\Jpsi$'s predominantly come from $g \to \chicJ$ fragmentation.}

	Without knowing the value for the nonperturbative matrix
element associated with $g \to c\cbar[{}^3S_1^{(8)}]$ fragmentation,
we cannot draw any definite conclusion regarding this possible $\psi'$
production mechanism.  But spin symmetry provides a powerful means for
testing the hypothesis that this proposal accounts for most of the
prompt $\psi'$'s observed at the Tevatron.  Recall that the
$^3S_1^{(8)}$ $c\cbar$ pair in \deltapsigraphs c is transverse up to
$O(M_{\psi'}^2/q_0^2)$ corrections.  This intermediate $j_\ell=0$
state transforms at leading order in the velocity expansion into a
$j'_\ell=0$ $\psi'$ through the $L=0$ emission or absorption of two
soft gluons.  The outgoing $\psi'$ in \deltapsigraphs c has exactly
the same angular momentum quantum numbers as the incoming hard gluon.
Therefore, $\psi'$'s produced at high transverse momenta via gluon
fragmentation to the final Fock state in eqn.~\psiwavefunc\ are 100\%
transversely aligned!

	This striking prediction can be tested by measuring the lepton
angular distribution in the $\psi' \to \ell^+ \ell^-$ decay channel.
The decay products are generally distributed according to
\eqn\psidecay{{d\Gamma \over d\cos\th}(\psi' \to \ell^+ \ell^-) \propto
(1 + \a \cos^2\th)}
where $\th$ denotes the angle between the lepton three-momentum in the
$\psi'$ rest frame and the $\psi'$ three-momentum in the lab frame.
The parameter $\a$ ranges over the interval $-1 \le \a \le 1$ and
equals unity for totally transverse $\psi'$'s.  Subleading corrections
to the infinite mass limit prediction will shift the value for $\a$ to
less than one.  But experimental observation of a lepton angular
distribution close to $d\Gamma/d\cos\th \propto (1 + \cos^2\th)$ would
provide strong support for this resolution to the $\psi'$ surplus
problem.

\bigskip\bigskip
\centerline{{\bf Acknowledgments}}
\bigskip

	We thank Eric Braaten and Sandip Trivedi for helpful discussions.

\listrefs
\listfigs
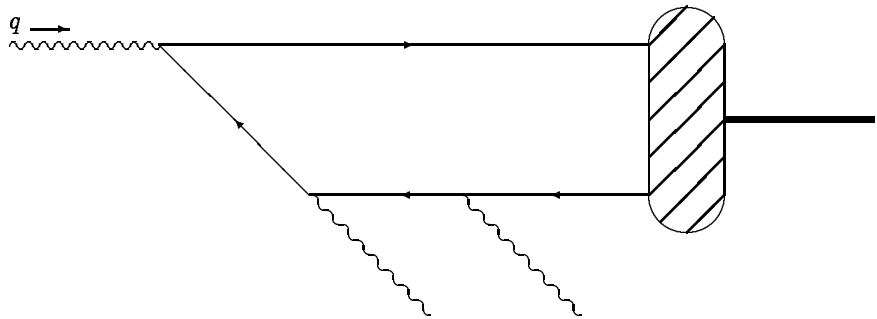
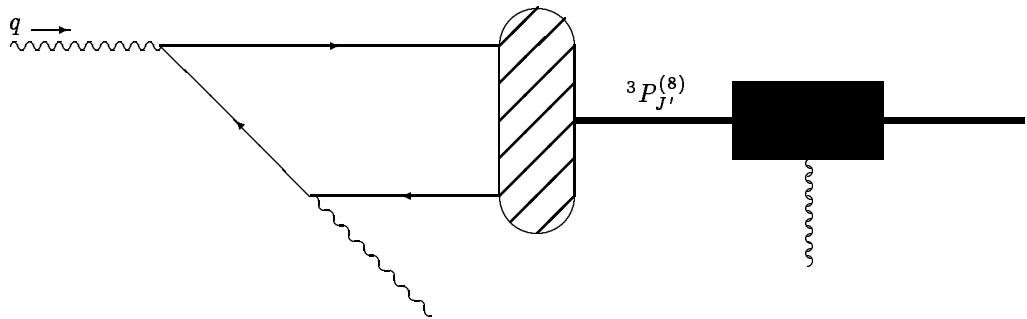
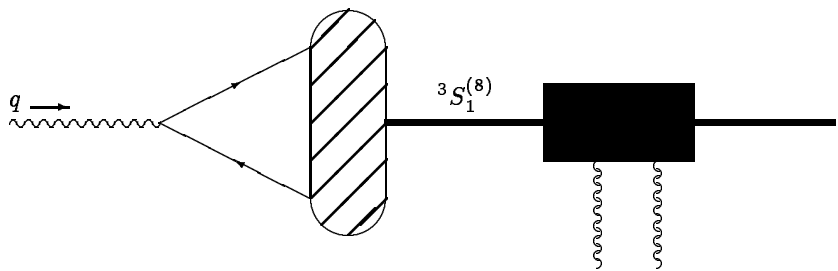
\bye